\documentclass{article}
\usepackage{amsmath, amssymb, amsfonts}
\title{ On a nonlinear gravitational wave. Geodesics } 
\author{Hristu Culetu, \\Ovidius University, Dept.of Physics and Electronics, \\B-dul Mamaia 124, 900527 Constanta, Romania, \\e-mail : hculetu@yahoo.com}

\begin{document}
\numberwithin{equation}{section}
\pagenumbering{arabic}
\maketitle
\newcommand{\fv}{\boldsymbol{f}}
\newcommand{\tv}{\boldsymbol{t}}
\newcommand{\gv}{\boldsymbol{g}}
\newcommand{\OV}{\boldsymbol{O}}
\newcommand{\wv}{\boldsymbol{w}}
\newcommand{\WV}{\boldsymbol{W}}
\newcommand{\NV}{\boldsymbol{N}}
\newcommand{\hv}{\boldsymbol{h}}
\newcommand{\yv}{\boldsymbol{y}}
\newcommand{\RE}{\textrm{Re}}
\newcommand{\IM}{\textrm{Im}}
\newcommand{\rot}{\textrm{rot}}
\newcommand{\dv}{\boldsymbol{d}}
\newcommand{\grad}{\textrm{grad}}
\newcommand{\Tr}{\textrm{Tr}}
\newcommand{\ua}{\uparrow}
\newcommand{\da}{\downarrow}
\newcommand{\ct}{\textrm{const}}
\newcommand{\xv}{\boldsymbol{x}}
\newcommand{\mv}{\boldsymbol{m}}
\newcommand{\rv}{\boldsymbol{r}}
\newcommand{\kv}{\boldsymbol{k}}
\newcommand{\VE}{\boldsymbol{V}}
\newcommand{\sv}{\boldsymbol{s}}
\newcommand{\RV}{\boldsymbol{R}}
\newcommand{\pv}{\boldsymbol{p}}
\newcommand{\PV}{\boldsymbol{P}}
\newcommand{\EV}{\boldsymbol{E}}
\newcommand{\DV}{\boldsymbol{D}}
\newcommand{\BV}{\boldsymbol{B}}
\newcommand{\HV}{\boldsymbol{H}}
\newcommand{\MV}{\boldsymbol{M}}
\newcommand{\be}{\begin{equation}}
\newcommand{\ee}{\end{equation}}
\newcommand{\ba}{\begin{eqnarray}}
\newcommand{\ea}{\end{eqnarray}}
\newcommand{\bq}{\begin{eqnarray*}}
\newcommand{\eq}{\end{eqnarray*}}
\newcommand{\pa}{\partial}
\newcommand{\f}{\frac}
\newcommand{\FV}{\boldsymbol{F}}
\newcommand{\ve}{\boldsymbol{v}}
\newcommand{\AV}{\boldsymbol{A}}
\newcommand{\jv}{\boldsymbol{j}}
\newcommand{\LV}{\boldsymbol{L}}
\newcommand{\SV}{\boldsymbol{S}}
\newcommand{\av}{\boldsymbol{a}}
\newcommand{\qv}{\boldsymbol{q}}
\newcommand{\QV}{\boldsymbol{Q}}
\newcommand{\ev}{\boldsymbol{e}}
\newcommand{\uv}{\boldsymbol{u}}
\newcommand{\KV}{\boldsymbol{K}}
\newcommand{\ro}{\boldsymbol{\rho}}
\newcommand{\si}{\boldsymbol{\sigma}}
\newcommand{\thv}{\boldsymbol{\theta}}
\newcommand{\bv}{\boldsymbol{b}}
\newcommand{\JV}{\boldsymbol{J}}
\newcommand{\nv}{\boldsymbol{n}}
\newcommand{\lv}{\boldsymbol{l}}
\newcommand{\om}{\boldsymbol{\omega}}
\newcommand{\Om}{\boldsymbol{\Omega}}
\newcommand{\Piv}{\boldsymbol{\Pi}}
\newcommand{\UV}{\boldsymbol{U}}
\newcommand{\iv}{\boldsymbol{i}}
\newcommand{\nuv}{\boldsymbol{\nu}}
\newcommand{\muv}{\boldsymbol{\mu}}
\newcommand{\lm}{\boldsymbol{\lambda}}
\newcommand{\Lm}{\boldsymbol{\Lambda}}
\newcommand{\opsi}{\overline{\psi}}
\renewcommand{\tan}{\textrm{tg}}
\renewcommand{\cot}{\textrm{ctg}}
\renewcommand{\sinh}{\textrm{sh}}
\renewcommand{\cosh}{\textrm{ch}}
\renewcommand{\tanh}{\textrm{th}}
\renewcommand{\coth}{\textrm{cth}}

\begin{abstract}
An exact, plane wave solution of the gravitational field equations is investigated. The source stress tensor is represented by an anisotropic null fluid with energy flux to which the energy density $\rho$ and all pressures are finite throughout the spacetime. They depend on a constant length (taken of the order of the Planck length) and acquire Planck values close to the null surface $t - z = 0$, the $z$-axis being the direction of propagation. However, $\rho$ and $p_{z}$ become positive when a cross-polarization term is introduced in the line-element. The timelike geodesics of a test particle are contained in a plane whose normal has constant direction and the null trajectories are comoving with a plane of fixed direction.
  \end{abstract}

\section{Introduction}
Gravitational waves (GWs) are ripples in the curvature of space and time which propagates at the speed of light. They are the mechanism by which the field of a changing mass distribution propagates, similar with the electromagnetic (EM) waves which represent the way by which a changing electric charge distribution tells remote observer that the source is varied \cite{LB}.

During their travel, the GWs squesh and stretch spacetime in the plane perpendicular to their direction of propagation. They however induce very tiny deformations: even very strong GWs from astrophysical sources are expected to perform relative length variations of order $10^{-21}$ \cite{EB}. Cabral and Lobo \cite{CL} showed that, in principle, the passage of a GW in a region with EM fields will have a measurable effect that may be computed studying Maxwell's equations on the perturbed background (the Minkowski space) of a GW. One obtains oscillations of the electric field which is aligned with the direction of the GW propagation. Even if in the absence of the GW the field was static, during the passage of the wave the electric field will become time-dependent, oscillating with the same frequency as the GW. Therefore, EM waves propagating along the same direction of the GW will be generated \cite{CL}.

Goswami et al. \cite{GMP} have shown that GWs propagate through ideal fluids without dissipation, but a nonzero shear viscosity will dissipate them. They calculate the dissipation of GW150914 \cite{BA} which propagated over a distance of 410 Mpc through the dissipative fluid and tested the data with the theoretical predictions. In their opinion, future observation of GWs at LIGO will give the possibility of detecting the viscosity of dark matter and dark energy.

As spacetime fluctuations that usually do not reflect from any matter surface, in the brane-world scenario zero-mode gravitons are trapped on a brane due to a nonlinear warping effect \cite{GM} and so Gogberashvili and Midodashvili expressed a radical possibility that GW151014 signal was caused by a short duration source in the brane-world scenario \cite{RS}. Liu et al. \cite{ML} check the Equivalence Principle (EP) about GW transient GW150914 associated with weak EM signal GW150914-GBM \cite{VC}, observed by Gamma-ray Burst Monitor. Their results show that the violation of the EP is quite insensitive to the location of the source. If the EP is right, initial gravitons will arrive the Earth (at Livingston and Hanford) in the same time. Caldwell, Devulder and Maksimova \cite{CDM} studied the general phenomenon of the conversion of a GW into a stationary gauge field. They show that GWs transform into tensor waves of a gauge field, similar with their conversion into an EM wave and back, when propagating through a stationary magnetic field. Lasky et al.\cite{PL} studied the so-called ''GW memory''- a permanent displacement of spacetime that comes from relativistic strong-field effects. They showed how the accumulation of several measurements will allow the detection of that effect.  

 In the weak gravitational field limit, one can expand Einstein's field equations $G_{ab} = 8\pi T_{ab}$ in powers of $h_{ab}$, with
  \begin{equation}
  g_{ab} = \eta_{ab} + h_{ab},~~~|h_{ab}| << 1
 \label{1.1}
 \end{equation} 
using a coordinate frame where (1.1) holds ($a, b$ run from 0 to 3, with $x^{0} = t,~ x^{1} = x,~ x^{2} = y,~x^{3} = z$) \cite{MTW, BS, EG, PG}. This is the linearized theory of gravity or tensor-field theory of gravity in flat spacetime. $h_{ab}$ in (1.1) is taken to be a perturbation of the flat spacetime caused, for example, by GWs, which are assumed to be plane waves at large distances from the source, as compared to their wavelength \cite{EG}. Let us mention that $h_{ab}$ is not a tensor. Nevertheless, it is convenient to think of a slightly curved spacetime as a flat spacetime with a ''tensor'' $h_{ab}$ defined on it \cite{BS}. One may therefore express, for instance, the curvature tensor $R^{a}_{~bcd}$ in terms of $h_{ab}$.

Poplawski \cite{NP} showed that real GWs may differ from solutions of the linearized field equations due to the fact that the two polarization modes are not independent and the monochromatic GW loses its periodic character because of the nonlinearity of Einstein's equations.

Pazouli and Tsagas \cite{PT} start from the fully nonlinear Einstein's equations before reducing them to their linear and second-order limits (around a chosen background), by employing the covariant $1+3$ splitting of the Weyl tensor for to describe the GWs instead of perturbing the background metric (see also \cite{RJS}). Their second order study suggest that Weyl-curvature distorsions could dictate the large-scale density perturbations. The key kinematical parameter is the shear tensor (in a $3+1$ covariant decomposition of the 4-velocity gradient) that is directly related to GWs in perturbed FLRW cosmologies.

  Usually, cosmologists take the linearized version of Einstein's equations for to deal with GWs. Even for the GW150914 event one assumes the linearized gravitational equations and implicitly the quadrupole formula may be used for the two BHs merger. In that case, strong gravity plays an important role and, therefore, the linear approximation should be used with caution. That is the main reason  which motivate us to work with the fully nonlinear Einstein's equations, to get the exact solutions and make approximations at a later stage of calculations. Therefore, we intend in this paper to follow Pazouli and Tsagas approach and to look for an exact solution of Einstein's equations that represents a nonlinear GW. We work firstly in Cartesian coordinates and seek a finite perturbation in spacetime, propagating with the velocity of light. We stress that that perturbation plays the role of a source of curvature and does not move in a given background. In section 2 we  compute the components of the source stress tensor in a spacetime which depends on a constant length, chosen to be of the order of the Planck length. The source fluid is anisotropic and can be expressed in terms of a null 4-vector and has negative energy density and pressure. Section 3 treats the same spacetime but in double-null coordinates $(v,~u)$ where the only nonzero component of the stress tensor is $T^{u}_{~v}$. Timelike and null geodesics are investigated in section 4. Finally, we discuss the implications of our results in section 5.

Throughout the paper we set $G = c = \hbar = 1$, unless otherwise specified.

\section{Nonlinear wave in Cartesian coordinates}  
Prior to look for a suitable spacetime of a nonlinear wave as an exact solution of the Einstein field equations, let us comment a little on the Eq. (1.1). Firstly we notice that the condition $|h_{ab}| << 1$ refers not to a weak field but to a weak (gravitational) potential. But what does ''1'' represent above? It is related to the background geometry, namely the Minkowski spacetime. However, Minkowski space is flat, which means no gravity. In other words, the ''weak'' perturbation $h_{ab}$ is compared to a background with zero gravity. Therefore, it turns out to be ambiguous to claim that a perturbation is small w.r.t. zero. The way out, in our opinion, is that the Minkowski space corresponds not to lack of gravity but to a constant gravitational potential, which is $c^{2} = 1$, where $c$ is the velocity of light in vacuo, so that a weak potential $h_{ab}$ means a potential much less than $c^{2}$. Put it differently, flat space concerns physics at constant gravitational potential, rendering it more Machian. That is valid even classically, i.e. the Newton potential at a distance $r$ from a point mass $m$ is not $-mG/r$ but $c^{2} - mG/r$. That will not change, of course, the equation of motion of a test particle but, however, it changes the sign of the potential.  

A similar situation arises when we consider ripples on the surface of a quiet lake. When we say ''small ripples'' we must add: w.r.t. what background? If the background means ''no ripples'' it is a nonsense to add the adjective ''small''. Therefore, the GWs are considered to be ''weak perturbations'' because from the very beginning they are taken to be ''weak''. So, it is not useless to look for strong GWs, to whom the linear approximation will no longer be valid. The fact that they are not observed experimentally is not a flaw. For example, Dark Energy and Dark Matter are not perturbations to some background and, nevertheless, they are still not observed experimentally. 

We look for an exact plane-wave solution of Einstein's equations in the form given by Misner, Thorne and Wheeler \cite{MTW} (see also \cite{BS})   
   \begin{equation}
  ds^{2} = - dt^{2} + \Psi^{2}(e^{-2\Phi} dx^{2} + e^{2\Phi} dy^{2}) + dz^{2}, 
 \label{2.1}
 \end{equation}
where $\Psi = \Psi(t-z),~\Phi = \Phi(t-z)$ are ''the background factor'' and ''the wave factor'', respectively. The linearized theory corresponds to $\Psi = 1$ and $\Phi << 1$. Then the metric (2.1) takes the form
   \begin{equation}
  ds^{2} = - dt^{2} + (1 - 2\Phi) dx^{2} + (1 + 2\Phi) dy^{2} + dz^{2}, 
 \label{2.2}
 \end{equation}
which is a plane wave propagating along the z - direction. It represents a perturbation of the Minkowski space. The line-element (2.2) is a solution of Einstein's field equations in the linear approximation in the so called (transverse-traceless) Lorentz gauge \cite{CL}. 

Motivated by our purpose to look for an exact solution of the fully nonlinear Einstein's equations, we introduce the following nonlinear gravitational wave 
   \begin{equation}
  ds^{2} = - dt^{2} + e^{-\frac{b}{\sqrt{(t - z)^{2} + b^{2}}}} dx^{2} + e^{\frac{b}{\sqrt{(t - z)^{2} + b^{2}}}} dy^{2} + dz^{2}, 
 \label{2.3}
 \end{equation}
where $b$ is a constant length that will be considered of the order of the Planck length. When $t - z >> b$, we may approximate $e^{-2\Phi} \approx 1 - 2\Phi,~ e^{2\Phi} \approx 1 + 2\Phi ,(\Phi = b/(2\sqrt{(t - z)^{2} + b^{2}})$) and (2.3) gives us (2.2), i.e. the linearized version. Our purpose now is to find what stress tensor $T_{ab}$ we need on the r.h.s of the Einstein equations in order (2.3) to be an exact solution. One finds that the only nonzero components of $T_{ab}$ are
   \begin{equation}
	8\pi \rho = 8\pi p_{z} = -8 \pi T^{t}_{~t} = 8 \pi T^{t}_{~z} = -8 \pi T^{z}_{~t} = -\frac{b^{2} (t - z)^{2}}{2[(t - z)^{2} + b^{2}]^{3}},
 \label{2.4}
 \end{equation}
where $\rho$ is the energy density of the anisotropic source, $p_{z} = T^{z}_{~z}$ is the pressure along the z-direction and $T^{t}_{~z}$ is the energy flux along the z-axis. It is worth noting that the source is a null fluid which can be put in the form \footnote{$l_{a}$ may be also expressed in terms of the null retarded coordinate $v = t - z$ as $l_{a} = -\partial_{a} v = (-1, 0, 0, 1)$}.
   \begin{equation}
	T_{ab} = \rho l_{a} l_{b},
 \label{2.5}
 \end{equation}
with $l_{a} l^{a} = 0,~l^{a} = u^{a} + s^{a},~u^{a} = (1, 0, 0, 0),~s^{a} = (0, 0, 0, 1),~l^{a} = (1, 0, 0, 1),~u_{a} s^{a} = 0,~u_{a} u^{a} = -1,~s_{a} s^{a} = 1,~l_{a} u^{a} = -1,~l_{a} s^{a} = 1$. $u^{a}$ here is the velocity vector field of a static observer ($x,y,z - const.$), $s^{a}$ is a spacelike vector orthogonal to $u^{a}$ and $l^{a}$ is a null vector related to the flow of energy on the z-direction. Comparing to the general form of an anisotropic fluid with energy flux \cite{HC}
  \begin{equation}
  T_{ab} = (p_{\bot} + \rho) u_{a} u_{b} + p_{\bot} g_{ab} + (p_{z} - p_{\bot}) s_{a}s_{b} +  u_{a} q_{b} + u_{b} q_{a},
 \label{2.6}
 \end{equation}
we see that 
  \begin{equation}
  p_{\bot} \equiv p_{x} = p_{y} = 0,~~~ q^{a} = - T^{a}_{~b}u^{b} - \rho u^{a}, 
 \label{2.7}
 \end{equation}
 where $q^{a} = (0, 0, 0, q^{z})$ is the energy flux 4-vector, with $q^{z} = T^{t}_{~z}$ the only nonzero component and $\sqrt{q^{a}q_{a}} = |\rho|$. The stress tensor may be also written in the form
  \begin{equation}
  T_{ab} = -\nabla_{a}\bar{\Phi} \nabla_{b}\bar{\Phi},~~~\bar{\Phi} = \Phi/\sqrt{4\pi}
 \label{2.8}
 \end{equation}
where $\nabla_{a}\bar{\Phi}$ is a null vector and $\bar{\Phi} = b/\sqrt{4\pi}(2\sqrt{(t - z)^{2} + b^{2}})$ could be interpreted as a scalar field with a ''wrong sign'' kinetic term \cite{HE,EH} and negative energy density (Ellis related his scalar field to a drainhole model but in Harris' approach the scalar field refers to a wormhole connecting two Reissner-Nordstrom (RN) black holes).

Let us observe that the energy density and pressure are negative but finite for any $t$ and $z$. They vanish at the null surface $t - z = 0$ and when $t - z \rightarrow \infty$. However, they acquire a minimum value $-1/108\pi b^{2}$ at $t - z = b/\sqrt{2}$. The density $\rho$ represents a lump of negative energy that moves on the z-direction with the speed of light. Far from the surface $t - z = 0$ (i.e. for $t - z >> b$) the energy density may be written as 
 \begin{equation}
 \rho = -\frac{\hbar c}{16\pi (t-z)^{4}},
 \label{2.9}
 \end{equation}
where we put $b = l_{P}$, with $l_{P}$ - the Planck length. Noting that in that region $\rho$ (and $p_{z}$) have a structure which remind us of the Casimir energy. We also have that the scalar curvature and the Kretschmann scalar are vanishing in the spacetime (2.3). 

How could we approach the null surface $t - z = 0$ , where the effects of the GW are stronger? To bring near that surface, we have to accelerate (of course, a huge acceleration is necessary), when the causal horizon $t - z = 0$ becomes a Rindler horizon for that accelerating observer \cite{MM}. The larger the acceleration, the closer to the null surface we are.

We consider now a congruence of static observers with the velocity vector field $u^{a} = (1, 0, 0, 0)$. All those observers are geodesic, namely the acceleration vector $a^{b} \equiv u^{a} \nabla_{a}u^{b} = 0$. In addition, the expansion scalar $\Theta \equiv  \nabla_{a}u^{a}$ also vanishes. However, the shear tensor 
\begin{equation}
\sigma_{ab} = \frac{1}{2}(h_{~b}^{c} \nabla_{c} u_{a}+ h_{~a}^{c} \nabla_{c} u_{b})-\frac{1}{3} \Theta h_{ab}+ \frac{1}{2}(a_{a} u_{b} + a_{b} u_{a}),
\label{2.10}
\end{equation}
where $h_{ab} = g_{ab} + u_{a}u_{b}$ is the projection tensor onto the direction perpendicular to $u_{a}$, has the nonzero components
 \begin{equation}
 \sigma^{x}_{~x} = - \sigma^{y}_{~y}= \frac{b(t - z)}{2[(t - z)^{2} + b^{2}]^{3/2}}.
 \label{2.11}
 \end{equation}
In other words, the gravitational wave distorts the spacetime in the plane $xOy$, orthogonal to the direction of propagation. We also notice that the distortion changes sign beyond the surface $t - z = 0$. 

We look now for the expression of the total energy flow $W$ measured by an observer laying at $z = z_{0} = const.$. It is given by \cite{HC}
 \begin{equation}
 W = \int{T^{a}_{~b}u^{b}n_{a}\sqrt{-\gamma}}dt~dx~dy
 \label{2.12}
 \end{equation}
 where $n^{a} = (0, 0, 0 , 1)$ is the unit normal to the surface of constant $z$, $\gamma_{ab} = g_{ab} - n_{a}n_{b}$ is the induced metric on that surface and $\gamma = det(\gamma_{ab}) = -1$. Eq. (2.12) yields
  \begin{equation}
 W = \frac{b^{2}}{16\pi}\int{\frac{(t - z_{0})^{2}}{[(t - z_{0})^{2} + b^{2}]^{3}}} dt~dx~dy
 \label{2.13}
 \end{equation}
If we change the variable of integration to $T = t - z_{0}$ and keeping in mind that 
  \begin{equation}
 \int{\frac{T^{2}}{[T^{2} + b^{2}]^{3}}dT} = \frac{T (T^{2} - b^{2})}{8b^{2}(T^{2} + b^{2})^{2}} + \frac{1}{8b^{3}} arctan\frac{T}{b},
 \label{2.14}
 \end{equation}
one obtains, after an integration from $0$ to $\infty$, that $W = (c^{4}/G) \Delta x \Delta y/256b$, where $\Delta x , \Delta y$ gives the domains of integration for $x$ and $y$, respectively.

\section{Nonlinear wave in double-null coordinates}
Given the flat form of the $t - z$ plane from the metric (2.3) and its dependence on its retarded null coordinate only, we find more practical to use the double null coordinates $v = t -z$ and $u = t + z$. The line-element (2.3) becomes now
 \begin{equation}
  ds^{2} = - dv du + e^{-\frac{b}{\sqrt{v^{2} + b^{2}}}} dx^{2} + e^{\frac{b}{\sqrt{v^{2} + b^{2}}}} dy^{2} , 
 \label{3.1}
 \end{equation} 
with $u$ - the advanced null coordinate. In this frame, the energy-momentum tensor of the source has a more simple form: there is only one nonzero component
 \begin{equation}
 8\pi T^{u}_{~v} = G^{u}_{~v} = \frac{b^{2}v^{2}}{(v^{2} + b^{2})^{3}},
 \label{3.2}
 \end{equation} 
i. e. the energy flux on the null direction $u$ (we take $v$ to play the role of time and the coordinates are $v,u,x,y$). The lump of gravitational energy travels with the speed of light and has a maximum $(T^{u}_{v})_{max} = 1/54\pi b^{2}$ at $v = b/\sqrt{2}$. One sees that  $T^{u}_{~v}$ vanishes on the surface $v = 0$ and very far away from that surface. It acquires there the form $T^{u}_{~v} \approx b^{2}/8\pi v^{4} = \hbar c/8\pi v^{4}$, which is a purely quantum quantity. 

In the spacetime (3.1) the source tensor is given by the same general form (2.6), but now we have $u_{a} u^{a} = -1$, $u^{a} = (1, 1, 0, 0)$, $s^{a} = (-1, 1, 0, 0)$ and $l^{a} = (0, 2, 0, 0)$. The energy flux vector has the components
   \begin{equation}
  q^{a} =  - T^{a}_{~b}u^{b} - \rho u^{a} = \left(\frac{1}{2}T^{u}_{~v},-\frac{1}{2}T^{u}_{~v}, 0,0\right).                                                                 
\label{3.3}
 \end{equation}
The energy density and pressures, rooted from (2.6) give now
   \begin{equation}
 \rho =  T_{ab}u^{a}u^{b} = -\frac{1}{2}T^{u}_{v} = -\frac{b^{2}v^{2}}{16\pi (v^{2} + b^{2})^{3}},~~~p_{u} = \rho,~~~p_{\bot} = 0,
 \label{3.4}
 \end{equation}
where $p_{\bot}$ represents the transversal pressures $p_{x}$, $p_{y}$ and $p_{u}$ is the pressure on the $u$-direction.  $q \equiv \sqrt{q^{a}q_{a}} = (1/2)T^{u}_{~v}$ and the acceleration of the congruence is again null, so that the congruence is geodesic. For example, taking $v = 10^{-15}s$ we have $q\approx b^{2}/16\pi v^{4} = 2\cdot 10^{-10} ergs/cm^{2}\cdot s$, a low value. It could be increased with a shorter value of the duration of measurement given by $v$. 

The two nonzero components of the shear tensor appears now as
 \begin{equation}
 \sigma^{x}_{~x} = - \sigma^{y}_{~y}= \frac{bv}{2(v^{2} + b^{2})^{3/2}}.
 \label{3.5}
 \end{equation}
We further write down the nonzero components of the Riemann tensor in the geometry (3.1). One obtains
 \begin{equation}
 \begin{split}
 R^{u}_{~xvx} = \frac{b}{2}\frac{bv^{2} - 2(2v^{2} - b^{2})\sqrt{v^{2} + b^{2}}}{(v^{2} + b^{2})^{3}} e^{-\frac{b}{\sqrt{v^{2} + b^{2}}}}\\
  R^{u}_{~yvy} = \frac{b}{2}\frac{bv^{2} + 2(2v^{2} - b^{2})\sqrt{v^{2} + b^{2}}}{(v^{2} + b^{2})^{3}} e^{\frac{b}{\sqrt{v^{2} + b^{2}}}}\\
   R^{x}_{~vvx} = \frac{b}{4}\frac{bv^{2} - 2(2v^{2} - b^{2})\sqrt{v^{2} + b^{2}}}{(v^{2} + b^{2})^{3}}\\
 R^{y}_{~vvy} = \frac{b}{4}\frac{bv^{2} + 2(2v^{2} - b^{2})\sqrt{v^{2} + b^{2}}}{(v^{2} + b^{2})^{3}}
 \label{3.6}
 \end{split}
 \end{equation}
It is worth noting that, close to $v = 0$, the above expressions behave as 
 \begin{equation}
 R^{u}_{~xvx} \approx  \frac{1}{eb^{2}},~~~R^{u}_{~yvy} \approx  -\frac{e}{b^{2}},~~~ R^{x}_{~vvx} =  -R^{y}_{~vvy} \approx  \frac{1}{2b^{2}},
 \label{3.7}
 \end{equation}
which are of the order of the Planck value. That takes place very near the null surface $v = 0$. However, in the region $v >> b$, the curvature tensor acquires negligible values
 \begin{equation}
 R^{u}_{~xvx} = -R^{u}_{~yvy} \approx  -\frac{2b}{v^{3}},~~~ R^{x}_{~vvx} =  -R^{y}_{~vvy} \approx  -\frac{b}{v^{3}},
 \label{3.8}
 \end{equation}
so that the spacetime becomes Minkowskian asymptotically.

\section{Geodesics}
To find the timelike geodesic equations in the spacetime (3.1), we start with the Lagrangean
  \begin{equation}
	L = \frac{1}{2} g_{ab} \dot{x}^{a} \dot{x}^{b} = -\dot{u} \dot{v} + \frac{1}{2}( e^{-\frac{b}{\sqrt{v^{2} + b^{2}}}} \dot{x}^{2} + e^{\frac{b}{\sqrt{v^{2} + b^{2}}}} \dot{y}^{2}),
 \label{4.1}
 \end{equation}
where a dot means a derivative w.r.t. the proper time $\tau$. If we keep track of the fact that the metric coefficient $g_{ab}$ do not depend on $u, x, y$, one obtains
  \begin{equation}
	  \dot{x} = C_{1} e^{\frac{b}{\sqrt{v^{2} + b^{2}}}},~~~ \dot{y} =  C_{2} e^{-\frac{b}{\sqrt{v^{2} + b^{2}}}},~~~ \dot{v} = C_{3},
 \label{4.2}
 \end{equation}
 where $C_{1}, C_{2}, C_{3}$ are dimensionless constants.
We take above $C_{3} = 1$ and so the variable $v$ will replace the proper time. The trajectories $x(v)$ and $y(v)$ may be obtained, in principle, from Eqs. 4.2, by integration. As far as the $u$ - coordinate is concerned, the function $u(v)$ may be obtained from the constraint $u_{a} u^{a} = -1$, where we have here $u^{a} = (\dot{v}, \dot{u}, \dot{x}, \dot{y})$. Therefore,
  \begin{equation}
	  \dot{u} = e^{-\frac{b}{\sqrt{v^{2} + b^{2}}}} \dot{x}^{2} + e^{\frac{b}{\sqrt{v^{2} + b^{2}}}} \dot{y}^{2} + 1,
 \label{4.3} 
 \end{equation}
where now $\dot{u} = du/dv$, etc. In order to satisfy that $\dot{x} < 1,~\dot{y} < 1$, we choose $C_{1} = 1/e$ and $C_{2} = 1$, to obtain
  \begin{equation}
	\dot{x}(v) = e^{\frac{b}{\sqrt{v^{2} + b^{2}}} -1},~~~\dot{y}(v) = e^{-\frac{b}{\sqrt{v^{2} + b^{2}}}}
 \label{4.4} 
 \end{equation}
While $\dot{y}(v)$ grows from $1/e$ at $v = 0$ to unity at $\infty$, $\dot{x}(v)$ drops from unity to $1/e$. They reach the same value $1/\sqrt{e}$ at $v = b\sqrt{3}$. Eq. (4.3) gives us now
  \begin{equation}
	  \dot{u}(v) = e^{\frac{b}{\sqrt{v^{2} + b^{2}}} -2} + e^{-\frac{b}{\sqrt{v^{2} + b^{2}}}} + 1
 \label{4.5} 
 \end{equation}
whence
  \begin{equation}
	  \dot{u}(v) = \frac{1}{e} \dot{x}(v) + \dot{y}(v) + 1,
 \label{4.6} 
 \end{equation}
or
  \begin{equation}
	  u(v) = \frac{1}{e} x(v) + y(v) + v,
 \label{4.7} 
 \end{equation}
where a constant of integration has been absorbed by $v$. In terms of the previous coordinates $(t,z)$ (4.7) yields
  \begin{equation}
	 x(t) + ey(t) -2ez(t) = 0
 \label{4.8} 
 \end{equation}
Eq. (4.8) shows that the trajectory of the test particle is contained in a plane to which $\vec{n}(1, e, -2e)$ is a normal vector. If the test particle has constant $x$ and $y$, say $x = x_{0},~y = y_{0}$, then $z = (x_{0} + ey_{0})/2e$ will also be constant, i.e. the particle will be at rest. That is a confirmation that the velocity field $u^{a} = (1, 0, 0, 0)$ corresponds to a geodesic.

To obtain the null geodesics we have only to get rid of the last term in Eq. (4.6). Hence
  \begin{equation}
	  \dot{u}(\lambda) = \frac{1}{e} \dot{x}(\lambda) + \dot{y}(\lambda),~~~\dot{v}(\lambda) = const.,
 \label{4.9} 
 \end{equation}
where $\lambda$ is the parameter along the null geodesic. But we have again $\dot{v} = const.$ so that $v = \lambda$ is a suitable choice. Passing from $(v,u)$ to Cartesian coordinates $(t,z)$, (4.9) yields
  \begin{equation}
	  \frac{1}{e} x(t) +  y(t) - z(t) = t,
 \label{4.10} 
 \end{equation}
which describes a plane moving with the speed of light along the direction given by its normal vector $\vec{n}(1/e, 1, -1)$. If the null test particle moves such that $x = x_{0},~y = y_{0}$, we get $z(t) = - t + (1/e)x_{0} + y_{0}$, which is a straight line.

\section{Nonlinear wave with cross-polarization}
In the linear version of the previous GW (Eq.2.2) the $+$ polarization mode
  \begin{equation}
	h_{xx} = -h_{yy} = -2\Phi = -\frac{b}{\sqrt{v^{2} + b^{2}}}.
 \label{5.1} 
 \end{equation}
 appeared. Let us introduce a $\times$ polarization through the last term of the metric
  \begin{equation}
	  ds^{2} = - dv du + e^{-\frac{b}{\sqrt{v^{2} + b^{2}}}} dx^{2} + e^{\frac{b}{\sqrt{v^{2} + b^{2}}}} dy^{2} + \frac{2b}{\sqrt{v^{2} + b^{2}}} dxdy 
 \label{5.2} 
 \end{equation}
where $h_{xy} = b/\sqrt{v^{2} + b^{2}}$. The source stress tensor gives us now an anisotropic fluid with energy flux and positive energy density
  \begin{equation}
	\rho (v) = p_{u}(v) = \frac{b^{2}}{4\pi (v^{2} + b^{2})^{2}},~~~q^{a} = (-\rho, \rho, 0, 0),~~~p_{\bot} = 0,~~~q = \rho,
 \label{5.3} 
 \end{equation}
with $\rho_{max} = \rho(0) = 1/4\pi b^{2}$. However, the scalar expansion
  \begin{equation}
	\Theta \equiv \nabla_{a}u^{a} = \frac{b^{2}}{v(v^{2} + b^{2})}
 \label{5.4} 
 \end{equation}
is nonzero and diverges at $v = 0$. We have again $u^{a} = (1, 1, 0, 0)$. Let us evaluate the order of magnitude of $\rho (v)$ from (5.3) for $v = 10^{-15}s$. With all fundamental constants inserted, one obtains $\rho = \hbar /c^{3}v^{4} = 3~ ergs/cm^{3}$ (the approximation $v >> b$ has been used). Here $v$ was taken as the duration of measurement. If the space were filled with these nonlinear GWs, to find a larger value of the energy density a very short-in-time measurement should be performed. We note that $\rho$ might be considered to be related to the dark energy. In addition, we have to keep in mind that the geometry (2.3) is not a vacuum solution of gravitational equations; the curvature is rooted from the gravitational wave itself, through the energy-momentum tensor (2.4).
The $u-v$ components of the shear tensor are given by
  \begin{equation}
	\sigma^{v}_{~v} = \sigma^{u}_{~u} = -\sigma^{u}_{~v} = -\sigma^{v}_{~u} = -\frac{b^{2}}{6v (v^{2} + b^{2})}
 \label{5.5} 
 \end{equation}
and
  \begin{equation}
	\begin{split}
	\sigma^{x}_{~x} =  \frac{b(b + 3\sqrt{v^{2} + b^{2}})}{6v (v^{2} + b^{2})},~~~\sigma^{y}_{~y} =  \frac{b(b - 3\sqrt{v^{2} + b^{2}})}{6v (v^{2} + b^{2})} \\
		\sigma^{x}_{~y} =  \frac{b(b - \sqrt{v^{2} + b^{2}})}{2v (v^{2} + b^{2})} e^{\frac{b}{\sqrt{v^{2} + b^{2}}}},~~~\sigma^{y}_{~x} = - \frac{b(b + \sqrt{v^{2} + b^{2}})}{2v (v^{2} + b^{2})} e^{\frac{-b}{\sqrt{v^{2} + b^{2}}}}												
\end{split}
 \end{equation}
It is worth noting that all nonzero components of the $\sigma^{a}_{~b}$ are vanishing when $v \rightarrow \infty$ and diverge at $v = 0$, excepting $\sigma^{x}_{~y}$ which vanishes at the origin.

\section{Conclusions}
We followed in this paper the Pazouli and Tsagas recipe and used the fully nonlinear Einstein equations, assuming a certain form of the starting geometry. Our ''perturbation'' solution plays the role of a source of curvature and does not act in a given background. The source stress tensor is given by a null fluid with no transversal pressures in Cartesian coordinates and with $\rho = p_{z} < 0$, which are very large near the null surface $t - z = 0$, with $z$ - the direction of propagation. We conjectured a new interpretation of the lack of gravitation in Minkowski space, concluding that flat space means constant Machian gravitational potential $c^{2}$. We also studied general timelike and null geodesics and found that they belong to a plane which has a constant direction, given by its normal vector. A static observer is not affected by the presence of the wave and a massless particle continues to move with the velocity of light.
We found also that the energy density $\rho$ becomes positive when the $\times$ polarization is included.\\

\textbf{Acknowledgement}\\
I am grateful to the anonymous referee for useful suggestions which improved the quality of the manuscript.

\end{document}